\documentstyle[prl,aps,multicol,epsf]{revtex}
\renewcommand{\narrowtext}{\begin{multicols}{2} \global\columnwidth20.5pc}
\renewcommand{\widetext}{\end{multicols} \global\columnwidth42.5pc} 

\begin{document}

\newcommand{\be}{\begin{equation}}
\newcommand{\ee}{\end{equation}}
\newcommand{\bea}{\begin{eqnarray}}
\newcommand{\eea}{\end{eqnarray}}
\newcommand{\nt}{\narrowtext}
\newcommand{\wt}{\widetext}

\title{Two different quasiparticle scattering rates 
in vortex line liquid phase of layered $d$-wave superconductors}

\author{D. V. Khveshchenko and A. G. Yashenkin$^{\dagger}$}

\address{Department of Physics and Astronomy, University of North
Carolina, Chapel Hill, NC 27599}
\maketitle

\begin{abstract}
We carry out a quantum mechanical analysis of the behavior of nodal quasiparticles 
in the vortex line liquid phase of planar $d$-wave superconductors. 
Applying a novel path integral technique we calculate a 
number of experimentally relevant observables and demonstrate that 
in the low-field regime the quasiparticle scattering rates deduced from 
photoemission and thermal transport data can be markedly 
different from that extracted from tunneling, specific heat, superfluid 
stiffness or spin-lattice relaxation time.
\end{abstract}

\nt
In recent years, the physics of nodal quasiparticles in planar $d$-wave 
superconductors such as the high-$T_c$ cuprates has attracted a lot of 
both, theoretical and experimental, attention. The spectroscopic and 
transport properties of these Dirac-like quasiparticles with a linear 
dispersion have been studied in quite some detail, and elaborate analyses 
of various mechanisms of elastic scattering in
the uniform superconducting state have been carried out, including 
the effects of potential \cite{Lee}, Kondo-like \cite{Kondo}, and extended
impurities \cite{Goldbart} as well as twin boundaries \cite{Durst}. 

In the mixed state, the recent quantum mechanical generalization of the 
earlier semiclassical approach \cite{Volovik} proposed in Ref.\cite{FT} allows one to account
for the non-uniformity of the local $d$-wave order parameter $\Delta_{\bf p}({\bf r})=
\Delta\cos(2\theta_{\bf p})\exp(i\phi({\bf r}))$ 
by means of a singular gauge transformation from the physical  
electrons $c_\sigma({\bf r})$ of spin $\sigma$ 
to the new fermionic quasiparticles. Unlike electrons,
the latter are subject to the effective magnetic field with zero mean 
which, besides the physical field, also includes the supercurrent circulating outside vortex cores. 

Applying the gauge transformation of Ref.\cite{FT}
to the electronic states with energies small compared to the maximum gap 
one can represent them in terms of the Nambu operators creating the 
auxiliary fermions with the momenta near the nodes of $\Delta_{\bf p}({\bf r})$
\bea
\pmatrix{c_\sigma({\bf r})\cr
 c^{\dagger}_{-\sigma}({\bf r}) } 
=\sum_{n=1,2;\pm} 
e^{\pm i{\bf k}_F^n{\bf r}} 
\pmatrix{e^{i\phi_A({\bf r})}u_{n\sigma}({\bf r})\cr
e^{-i\phi_B({\bf r})}v_{n\sigma}({\bf r}) },
\eea
where $n=1,2$ labels the pairs of the opposite nodes,
while the choice of the phases $\phi_{A,B}({\bf r})$ is only restricted 
by the condition $\phi_A({\bf r})+\phi_B({\bf r})=\phi({\bf r})$.
 
In the case of a regular vortex lattice, the representation (1)
was used to demonstrate that the structure of the quasiparticle energy 
spectrum is that of the energy bands, rather than the Landau levels
\cite{Halperin}. 

In the present paper, we extend the analysis based on the representation (1) 
to the experimentally well-documented vortex line liquid (VLL) phase  \cite{review}
where the vortices are distributed totally randomly due to their strong pinning by 
columnar or other defects. 

Any disorder, including that induced by random vortices, is expected to 
affect the behavior of the $d$-wave quasiparticles most strongly at the 
lowest energies, possibly resulting in the complete quasiparticle
localization which, however, is still awaiting for experimental confirmation \cite{Toronto}.
In what follows, we focus on the ballistic regime of quasiparticle energies 
large compared to the localization scale (see below) which is readily accessible 
by a number of standard probes such as angular-resolved photoemission (ARPES), 
thermal transport, tunneling, specific heat, muon spin rotation ($\mu$SR), and 
spin-lattice relaxation.

In the quantum-mechanical approach of Ref.\cite{FT}, 
the combined effect of external magnetic field ${\bf H}=\nabla\times{\bf A}({\bf r})$ 
and swirling supercurrent characterized by the superfluid 
velocity ${\bf v}_s({\bf r})=\frac{\hbar}{2}(\nabla\phi_A+\nabla\phi_B)-{e\over c}{\bf A}$ 
does not amount solely to the semiclassical Doppler shift 
$\epsilon\to \epsilon-{\bf k}_F{\bf v}_s({\bf r})$ of the quasiparticle 
energies \cite{Volovik}. 
The latter is to be complemented by the vector potential, 
${\bf a}({\bf r})=\frac{\hbar}{2}(\nabla\phi_A-\nabla\phi_B)$, 
that couples to the quasiparticles via their momentum, 
${\bf k}\to {\bf k}-{\bf a}({\bf r})$,
and accounts for the quantum mechanical Berry phase corresponding to their 
Bohm-Aharonov (BA) scattering by the vortices. Thus, the complete Hamiltonian 
of the non-interacting nodal quasiparticles contains 
both the scalar- and the vector-like random terms 
\be
{\cal H}=\sum_{n,\sigma}\int d{\bf r}\,
{\overline \psi}_{\sigma}[{\hat \gamma}_0{\bf k}^n_F{\bf v}_s({\bf r})+
v_i{\hat {\gamma}}_i({p}_i-{a}_i({\bf r}))]\psi_\sigma.
\ee
In Eq.(2), we used the $4\times 4$ representation for the matrices 
${\hat \gamma}_\mu=({\hat \sigma}_2, i{\hat \sigma}_1, i
{\hat \sigma}_3)\otimes{\hat \sigma}_3$ acting 
in the space of the Dirac bi-spinors composed of the Nambu spinors: 
$\psi_\sigma=[(u_{1\sigma}, \epsilon_{\sigma\sigma^\prime}v_{1\sigma^\prime}),
(u_{2\sigma}, \epsilon_{\sigma\sigma^\prime}v_{2\sigma^\prime})
({\hat \sigma}_1+{\hat \sigma}_3)/{\sqrt 2}]$. 

In order to keep our discussion and formulas relatively simple 
we consider the case of isotropic quasiparticle dispersion and use the units where $v_i=hc/2e=k_B=1$.
Moreover, because of the predominantly small-angle nature of the quasiparticle 
scattering by the vortices we choose to neglect the 
processes of inter-node scattering. While anticipating that 
neither of these simplifying assumptions will affect our main conclusions,
we recognize that a combination of the above factors in the case of the real cuprates 
may give rise to an additional one-to-two dimensional crossover regime \cite{1D}.

As in the previous studies of the VLL phase \cite{Franz}, 
we average over different vortex configurations 
by assuming the Gaussian distributions 
\be
\langle{\rm g}_{(v,a)}^i({\bf q}){\rm g}^j_{(v,a)}(-{\bf q})\rangle=
w_{(v,a)}({\bf q})\left( \delta_{ij}- {{q}_{i} 
{q}_j\over {\bf q}^2} \right),
\ee
for both ${\bf g}_v={\bf v}_s$ and ${\bf g}_a={\bf a}$.
The striking difference between $w_v({\bf q})={\alpha/({\bf q}^2+\alpha)}$ 
and $w_a({\bf q})={\alpha/{\bf q}^2}$
which are both proportional 
to the areal density of vortices $\alpha=2\pi H$ 
reflects the presence of screening for the scalar-like Doppler potential
and its absence for the vector-like BA scattering.
The longitudinal ($\propto {q}_i {q}_j/{\bf q}^2$)
part of the correlator (3) proves to contribute negligibly to all the quantities of interest
(besides, it is suppressed by the Coulomb interactions). 

In contrast to the previous analyses which focused solely on the effect
of Doppler scattering \cite{Franz}, 
we find that vortex disorder has a profound effect on the 
quasiparticle spectrum which can not be adequately modelled by a constant
quasiparticle width. In order to illustrate this 
point, we apply the standard self-consistent Born equation  
\be
\Sigma_{}(\epsilon,{\bf p})=\int {d{\bf q}\over (2\pi)^2}
{\epsilon+\Sigma_{}(\epsilon,{\bf q})
\over {\bf q}^2-(\epsilon+\Sigma_{}(\epsilon,{\bf q}))^2}
w_{(v,a)}({\bf p}+{\bf q})
\ee
to the separate contributions of the two scattering
mechanisms towards the total quasiparticle width
${\rm Im}\Sigma_v+{\rm Im}\Sigma_a$.

In the case of scalar disorder Eq.(4) yields 
${\rm Im}\Sigma_v\propto\alpha^{1/2}$ for small energies and momenta ($\epsilon,
p\lesssim\alpha^{1/2}$) 
while at ${\rm max}(\epsilon, p)\gg\alpha^{1/2}$ it behaves as 
$\propto\alpha/{\rm max}(\epsilon, p)$ which dominates 
over the scattering by the vortex cores
whose rate is estimated as $\propto \alpha/\Delta$.

At first sight, the effect of the BA scattering may seem to be much stronger,
since a naive solution $\Sigma_a$ of Eq.(4) with the singular 
kernel $w_a({\bf q})$ is plagued with a logarithmic 
infrared divergence of the momentum integral.
 
In order to avoid this spurious divergence which stems from 
the non-gauge invariant nature of the auxiliary fermion propagator 
one has to proceed directly with computing the manifestly 
gauge-invariant (retarded) Green function of the physical electrons 
$$
G^R(\epsilon,{\bf r})=\int^\infty_0 dt e^{i\epsilon t}\langle 
c(t,{\bf r})c^\dagger(0,{\bf 0})\rangle=\sum_{n=1,2;\pm} e^{\pm i{\bf k}_F^n{\bf r}}
$$
\be
\langle\psi_n({\bf r})\,\exp[-i\int_{C}({\bf v}_s+{\bf a}\,\sigma_3
\otimes{\bf 1}) d{\bf r}^\prime]\, {\overline \psi}_n({\bf 0})\rangle.
\ee
It turns out that the exponential decay of this function (see below) makes it
largely independent of the contour $C$ which can then be chosen as the 
straight path between the end points $\bf r$ and $\bf 0$.

To compute the amplitude (5) we apply the path-integral
representation of Ref.\cite{Stefanis} to the propagator of the auxiliary Dirac fermions.
First, for a fixed vortex configuration, we cast Eq.(5)  
in the form of a functional integral over the space-time coordinate 
${\bf r}(\tau)$ and the conjugate momentum ${\bf p}(\tau)$ parameterized by the 
proper time $\tau$ 
$$
{\cal G}^R(\epsilon,{\bf r}\,|\,{\bf v}_s, {\bf a})=
\int^\infty_0 d\tau 
\int_{{\bf r}(0)=0}^{{\bf r}(\tau)={\bf r}}
D{\bf r}\,D{\bf p}\,e^{i{\hat S}_0[\tau]}
$$
\be
\exp[i\int^\tau_0 d\tau^\prime({\bf k}_F{\bf v}_s+
{d{\bf r}\over d\tau}{\bf a})-i\int_{C}({\bf v}_s+{\bf a}) d{\bf r}^\prime)],
\ee
where we dropped, for the sake of 
compactness, the sum over the nodal points and introduced the free fermion action
\be 
{\hat S}_0[\tau]=\int^\tau_0 d\tau^\prime 
[\epsilon{\hat \gamma}_0+{\bf p}({d{\bf r}\over d\tau^\prime}-{\hat {\bf
\gamma}})].
\ee
Averaging over the disorder variables ${\bf v}_s $ and $ {\bf a} $
with the use of Eq.(3) results in the electron Green function 
$ G^R (\epsilon, {\bf r})= \langle {\cal G}^R (\epsilon, {\bf r}|{\bf v}_s, 
{\bf a}) \rangle $ which is given by Eq.(6) where, instead of the exponential 
phase factor, the integrand contains a product of two attenuation factors 
$$
W_{(v,a)}[{\bf r}(\tau)]=\exp[-{1\over 2}\int {d{\bf q}\over (2\pi)^2}
\int^\tau_0 d\tau_1\int^\tau_0d\tau_2
$$
\be 
{\bf u}_i(\tau_1){\bf u}_i(\tau_2)w_{(v,a)}({\bf q})e^{i{\bf q}
({\bf r}(\tau_1)-{\bf r}(\tau_2))}],
\ee
with $ {\bf u}_v={\bf v}_F-{\bf r}/\tau $ and 
${\bf u}_a=d{\bf r}/d\tau -{\bf r}/\tau $.
Thus, the presence of the exponent of the line interal 
taken along the contour $C$ in Eq.(5) strongly reduces 
the effect of both ${\bf a}$ and ${\bf v}_s$, as compared to the case of the 
gauge-variant propagator of the auxiliary fermions.
This observation seems to have been overlooked in the 
earlier studies of this and related problems where the phase factor in 
question would either not appear at all \cite{Millis} or be averaged 
separately from the fermion propagator computed in a different approximation \cite{Dorsey}.

Proceeding along the lines
of the previous analyses of the problem of non-relativistic 
fermions subject to a random vector potential \cite{RMF}, one can show that 
in the ballistic regime (which in the present case is defined by the condition $\epsilon\gg\alpha^{1/2}$)
the path integral (6) is dominated by the fermion trajectories which only slightly 
depart from the straight line ${\bf r}_0(\tau)={\bf v}\tau$. 
Evaluating the factor $W_v$ for such a trajectory, one obtains 
\be
W_v[{\bf r}(\tau)]
\approx \exp\left[-r\left({\sqrt {\tau\over r}}-{\sqrt {r\over \tau}}\right)^2
\int^\infty_0 {d{q}\over 2\pi}w_v({q})\right],
\ee
while the integral in $W_a$ turns out to be 
proportional to the so-called Amperian
area of the closed contour composed of a fermion trajectory 
${\bf r}(\tau)$ and the "return" path $-{\bf r}_0(\tau)$.
Although this purely geometrical term vanishes for the
saddle-point trajectory ${\bf r}(\tau)={\bf r}_0(\tau)$, 
its expansion to first order in the transverse deviation $r_\perp(\tau)$ yields   
\be
W_a[{\bf r}(\tau)]\approx\exp[-{\alpha\over 2}\int^\tau_0 
d\tau^\prime |r_\perp(\tau^\prime)|].
\ee
By analogy with the non-relativistic problem studied
in Refs.\cite{RMF} the path integral (6) with the $W_i$ factors given by Eqs.(9) and (10) can 
be related to the resolvent of the Schroedinger equation 
describing the transverse motion of the Dirac fermion 
$$
\left[\partial^{2}_{x} + (\epsilon^2-q^2) +
\left(\alpha\, \epsilon \, /q\right)^2\left(|x|+x_0)\right)^2+
i\alpha \, {\rm sign}\, x \right]
$$
\be
\times
g(\epsilon, q|x, x^\prime)=\delta(x-x^\prime),
\ee
where $x_0=( {\sqrt {|\epsilon/q|}}-{\sqrt {|q/\epsilon|}} )^2/2\alpha^{1/2}$. 

By analogy with the results of Refs.\cite{RMF}
the averaged physical electron propagator ${G}^R(\epsilon,{\bf p})$ can be obtained
by convoluting the kernel $1/({\bf p}^2-q^2)^{3/2}$ 
with the solution of (11) taken at $x=x^\prime=0$ which is given by the formula  
\be
g(\epsilon, q|0, 0)=\left[{d\over dx}\ln(U_+U_-)\right]^{-1}_{x=0}
\ee
where $U_{\pm}=U\left(a_{\pm},\lambda [x_0\pm x]\right)$ is the parabolic cylinder function  
of the parameter $a_{\pm}=(\epsilon^2-q^2\pm i\alpha)/\lambda^2$ 
and $\lambda=({2i\alpha |\epsilon/q |})^{1/2}$.

Turning now to the applications of Eq.(5), we first discuss the electron spectral 
function satisfying the dispersion relation
$G^R(\epsilon,{\bf p})=\int A(\epsilon^\prime,{\bf p})d\epsilon^\prime/
\pi(\epsilon-\epsilon^\prime+i\delta)$.
Near the maximum, $|\epsilon^2-{\bf p}^2|\lesssim \alpha$, it can be shown
to take the form
\be
A(\epsilon,{\bf p})\approx{(\epsilon{\hat \gamma}_0-{\bf p}{\hat {\bf 
\gamma}}){\sqrt \beta}
\over {[(\epsilon^2-p^2)^2+\beta^2(1+(|\epsilon/p|-1)^4/4)]^{3/4}}}
\ee
where $\beta\approx{\pi}\alpha/8$, thereby demonstrating a replacement
of the bare pole by a branch cut of the function $(z-z_0)^{3/2}$ resulting from 
the above convolution procedure. 
According to Eq.(13), the decay of the electron propagator in real space
($G^R(\epsilon,{\bf r})\propto \epsilon(\Gamma/r)^{1/2}
e^{-\Gamma r}$ for $r\gg 1/\Gamma$) is governed by
$\Gamma(\epsilon)\propto {\alpha/\epsilon}$ 
which should be thought of as the actual (energy-dependent) quasiparticle width.

This direct experimenta prediction can be tested by performing ARPES measurements 
in the VLL phase of the cuprates under the weak-field conditions 
(${\sqrt H}\ll T\ll\Delta$).

Also, comparing Eq.(13) with the estimate for ${\rm Im}\Sigma_v$ obtained from 
Eq.(4), we conclude that in the ballistic regime both the BA and the Doppler 
scattering mechanisms appear to be equally important, contrary to the conclusions drawn 
in Ref.\cite{Ye1}.

Next, we compute thermal conductivity given by the averaged product of two 
electron propagators 
\be
\kappa_{xx}=\int^\infty_0 {(\epsilon/T)^2d\epsilon\over \cosh^2(\epsilon/2T)}
\int {d{\bf r}\over 2\pi}
{\rm Tr}
\langle{\hat {\gamma}}_1{\cal G}^A(\epsilon,{\bf r})
{\hat {\gamma}}_1{\cal G}^R(\epsilon,-{\bf r})\rangle.
\ee
The corresponding path integral reads as  
$$
\langle
{\hat {\bf \gamma}}{\cal G}^A(\epsilon,{\bf r}){\hat {\bf \gamma}}
{\cal G}^R(\epsilon,-{\bf r})\rangle =
\int^\infty_0 d\tau_1 d\tau_2 
\int_{{\bf r}_{1,2}(0)=0}^{{\bf r}_{1,2}(\tau_{1,2})=\pm{\bf r}}
$$
\be
\prod_{\alpha,\beta=1,2}D{\bf r}_\alpha D{\bf p}_\beta 
{\hat {\bf \gamma}}e^{i{\hat S}_0[\tau_1]}{\hat {\bf \gamma}}
e^{i{\hat S}_0[\tau_2]}\prod_{i=a,v}W_{i}({\bf r}_\alpha-{\bf r}_\beta),
\ee
where the factors $W_i({\bf r}_\alpha-{\bf r}_\beta)$ with 
$\alpha\neq\beta$ account for the vertex corrections, alongside the 
self-energy ones ($\alpha=\beta$). Thus, the path-integral method 
of computing the Dirac fermion conductivity is capable of proceeding beyond the 
conventional (non-crossing and fan-shaped) ladder series of the vertex 
corrections to the bare fermion bubble (the latter suffice only  
if the number $N$ of the Dirac species is large, while in the $d$-wave problem $N=\sum_\sigma 1=2$).

Upon integrating over the "center of mass" variables ${\bf r}_1+{\bf r}_2$
and ${\bf p}_1+{\bf p}_2$ and rescaling the ones describing
relative motion, we again arrive at Eq.(11). This time around, 
it is formulated in terms of the transverse relative coordinate
$(r_{1\perp}-r_{2\perp})$, and its solution yields 
the average $\langle {\cal G}^A{\cal G}^R\rangle\propto
\epsilon(\Gamma_{tr}/r)^{1/2}e^{-\Gamma_{tr}r}$, where $\Gamma_{tr}=2\Gamma$
is now playing the role of the momentum relaxation rate, in agreement with the above 
estimate for the quasiparticle spectral width. Plugging this asymptote into
Eq.(14) we find that in the low-field regime the thermal conductivity behaves as 
\be
\kappa_{xx}(T,H)\simeq {7\pi^2\over 15}{T^3\over H}.
\ee
While being in agreement with the estimate based on the kinetic equation 
$\kappa\propto T\nu/\Gamma_{tr}$ proportional to the linear density of states 
(DOS) $\nu\propto T$ and $\Gamma_{tr}\propto H/T$, Eq.(16) is strikingly different from 
the result ($\kappa\propto T^2/H^{1/2}$) that one would  
obtain by naively assuming that the 
quasiparticle width remains constant ($\Gamma^{\prime} \propto H^{1/2}$) up to
the energies $\epsilon\sim T$ (cf. with Refs.\cite{Franz}).

The analysis of the data of Ref.\cite{Ong} taken in $YBa_2Cu_3O_{6.99}$ 
shows that Eq.(16) should be expected to hold for $T<30K$ and $0.1\lesssim{\sqrt H}/T\lesssim 1$
where the vortex-induced $\Gamma_{tr}$ dominates over the other 
mechanisms of scattering, including potential
impurities. 

Our approach also enables one to compute 
other averages, such as $\langle {\cal G}^R{\cal G}^R\rangle\propto\epsilon
(\Gamma_{tr}/r)^{1/2}e^{2i\epsilon r-\Gamma_{tr}r}$, which controls 
the effect of vortex disorder on superfluid stiffness measured by $\mu$SR   
\be
{\rho_s(0)-\rho_s(T)}=
{1\over \pi k_F}{\rm Im}\int {{d\epsilon}\tanh\left({\epsilon\over 2T}\right)}
\ee
$$
\int d{\bf r}{\rm Tr}\langle {\cal G}^R(\epsilon,{\bf r}){\cal G}^R(\epsilon,-{\bf r})\rangle
\simeq {2\ln 2\over \pi}T+{H\over 8\pi T}\ln \frac{T^2}{H},
$$
Notably, in contrast to the spectral and transport characteristics
whose behavior is determined by the structure of the electron spectral function near its maximum, 
Eq.(17) is governed by the overall momentum integral of the solution of the two-particle analog of Eq.(11). 

This behavior is common amongst the thermodynamic quantities 
associated with the averages of local bi-linear operators 
${\overline \psi}_\sigma({\bf r}) \psi_{\sigma^{\prime}}({\bf r})$ which are invariant
under the gauge transformation (1). In fact, such averages as, e.g., 
$G^R(\epsilon,{\bf 0})$ must be computed differently, since now the 
trajectories contributing to the path integral (6) may deviate very strongly 
from the semiclassical one, ${\bf r}_0(\tau)={\bf 0}$, although the mean 
square of the distance over which a typical trajectory ${\bf r}(\tau)$ ventures 
from the origin still scales quadratically with time, 
$\langle{\bf r}^2(\tau)\rangle\propto \tau^2$. For $1/\Delta\lesssim\tau$ 
this allows one to evaluate the damping factors as  
\be
W_i[{\bf r}(\tau)]\approx\exp\left[-{\tau^2\over 2}\int {d{\bf q}\over
(2\pi)^2}w_i({\bf q})\right].
\ee
Plugging (18) instead of (8) into (6) and computing the resulting (quadratic) 
path-integral, we arrive at the correction to the linear 
DOS corresponding to the clean limit 
\be
\nu(\epsilon)={\rm Im} {\rm Tr} \left[{\hat \gamma}_0{G}^R(\epsilon,{\bf 0})
\right]= {\sigma_H\over \pi^{3/2}}{\cal F}\left({\epsilon\over \sigma_H}\right),
\ee 
where $\sigma^2_H\propto H\ln(\Delta^2/H)$ and ${\cal
F}(x)=\pi^{1/2}x \,{\rm Erf}(x)+\exp(-x^2)$.
The effect of vortex disorder is most pronounced at small energies $\epsilon\lesssim H^{1/2}$, 
and it appears to be stronger than in the semiclassical (Doppler-only) approximation (cf. with \cite{Vekhter}).  

Directly, this DOS correction can be extracted from the tunneling conductance $G(V)\propto\nu(V)$.
Indirectly, it can also be manifested through the correction to electronic specific heat    
\be
C(T)=\int^\infty_0 {(\epsilon/T)^2\nu(\epsilon)d\epsilon\over 
\cosh^2(\epsilon/2T)}\simeq
{18\zeta(3)T^2\over \pi}+{\sigma_H^4\over 16\pi T^2}
\ee
Notably, the correction $\Delta C\propto H^2\ln(\Delta^2/H)/T^2$ is 
smaller than the result ($\Delta C\propto H$) obtained under in the situation where the inter-vortex
repulsion is stronger than random pinning, and therefore the VLL is partially ordered \cite{Vekhter}.

Although the smallness of the disorder-induced term in Eq.(20) might hinder its
detection, an alternate possibility is offered by the spin-lattice relaxation time 
\be
{1\over T_1(T)}\propto\int^\infty_0 {\nu^2(\epsilon)d\epsilon\over 
\cosh^2(\epsilon/2T)}\simeq
{2\over 3}T^3+{\sigma_H^{3} \over 3\pi^{5/2}},
\ee 
where we dropped the overall prefactor proportional 
to the ion-specific
matrix elements.

In summary, we carried out a fully quantum mechnical 
analysis of the quasiparticle properties of the VLL phase 
of layered $d$-wave superconductors. We demonstrated that both the semiclassical Doppler shift 
and the intrinsically quantum mechanical BA scattering have comparable effects on 
all the observables. Our path-integral approach enabled us 
to identify the energy-dependent effective quasiparticle width $\Gamma(\epsilon)\propto H/\epsilon$ 
describing the near-maximum ($\epsilon^2\approx {\bf p}^2$)
behavior of the distinctly non-Lorentzian electron spectral function
which can be directly measured by ARPES.

Also, we exposed the striking difference between the two rates:   
$\Gamma_{tr}\sim\Gamma(T)\propto H/T$ displayed by the transport characteristics  
and $\Gamma^\prime\sim\Gamma(H^{1/2})\propto H^{1/2}$ manifested by tunneling and thermodynamic 
quantities.

In contrast, the real-space averaging procedure applied in the previous 
studies of the disordered vortex states \cite{Vekhter} is bound to deliver the latter rate,
because, focusing solely on the energy distribution, it does not  
faithfully represent the momentum dispersion of the averaged electron spectral function.

We emphasize that the origin (that is, a non-trivial energy dependence)
of the different apparent 
quasiparticle rates, as revealed by the different measurements, must be 
distinguished from the conventional juxtaposition 
of quasiparticle lifetime versus transport time in, e.g., photoemission and transport experiments.

We expect that, albeit derived under a number of simplifying assumptions, our 
main conclusions are robust against including 
such factors as spatial anisotropy and non-linearity of the fermion dispersion as
well as inter-node scattering and, most importantly, 
modifying the distributions (3) with the purpose of describing 
partially ordered "vortex glass" states \cite{Vekhter}.

Lastly, the results of this paper can also be used to describe the effects of thermal
phase fluctuations controlled by $\alpha\propto T$ in the pseudogap phase of the cuprates \cite{Millis,Dorsey}.
To this end, we predict that different measurements 
may return different values of the effective quasiparticle width, which might 
explain the inconsistency between the widths deduced from the tunneling and ARPES data \cite{Millis}.
 
This research was supported by the NSF under Grant No. DMR-0071362.

\wt
\end{document}